\documentclass[twocolumn, twocolappendix]{aastex63}

\usepackage{amsmath}
\usepackage{graphicx}

\newcommand{\HL}{\mathcal{H}_L}
\newcommand{\HU}{\mathcal{H}_U}

\begin{document}

\title{Identifying strong gravitational-wave lensing during the second observing run of Advanced LIGO and Advanced Virgo}

\author{Xiaoshu Liu}
 \email{xiaoshu@uwm.edu}
\author{Ignacio Maga\~na Hernandez}
 \email{maganah2@uwm.edu}
\author{Jolien Creighton}
 \email{jolien@uwm.edu}
\affiliation{
 Department of Physics, University of Wisconsin-Milwaukee, Milwaukee, WI 53201, USA
}

\begin{abstract}

We perform Bayesian model selection with parameter estimation to identify potentially lensed gravitational-wave images from the second observing run (O2) of Advanced LIGO and Advanced Virgo. Specifically, we compute the Bayesian evidence for a pair of events being lensed or not lensed (unlensed) using nested sampling. We consider the discrete coalescence phase shifts that can be induced if the gravitational-wave signal interacts with the lens caustics in the model selection. We find that the pair of events, GW170104 and GW170814 with a $\pi/2$ coalescence phase shift, has a significant Bayes factor ($B^L_{U}$ $\sim 1.98 \times 10^4$) favoring the lensing hypothesis. However, after taking into account the long time delay of approximately 7 months between events, the timing Bayes factor is significantly small ($B_t \sim 8.7\times10^{-2}$). The prior probability for detecting strongly lensed pairs at O2 sensitivity are exceedingly small for both galaxy and galaxy cluster lensing. Combining the lensing and timing Bayes factors with the prior odds on lensing gives an odds ratio of $O^L_{U} \sim 20$. However, the model dependence of the timing and prior odds factors does not provide strong evidence to demonstrate that the pair is strongly lensed.
\end{abstract}


\section{\label{sec:level1}Introduction}

When gravitational waves (GW) propagate near massive galaxies or galaxy clusters, similar to light, the GW can be strongly lensed. If the massive galaxies or galaxy clusters are along the line of sight of the GW source, gravitational-wave observatories are expected to see multiple images with a time delay of hours to weeks \citep{Haris:2018arXiv180707062H} from the same astrophysical source as long as both images are above the GW detection threshold. 
Based  on predictions  on  the  number  of  expected GW sources, and the distribution of lenses in the Universe, \citep{Li:2018prc, Oguri:2018muv} suggests that around one in a thousand events observed by Advanced LIGO and Advanced Virgo \citep{{Aasi:2013wya}, {TheLIGOScientific:2014jea}, {Harry:2010zz}, {TheVirgo:2014hva}} at design sensitivity will be lensed. The rate computations typically assume that a single image is detected, and the majority of the lenses are galaxy lenses. The lensing rate is expected to be lower at O2 sensitivity and even lower when considering double images. Galaxy cluster lenses have been investigated in \citep{Smith:2017mqu,Smith:2018gle,Smith:2018kbc,Dai:2020tpj,Robertson:2020mfh}, which find that the rate of galaxy cluster lensing is around $10^{-5} \, \rm yr^{-1}$ at O1 sensitivity, and given the fact that the small sensitivity improvement in O2, we do not expect the rate has notably increased. Lensing event rates can also be inferred with the measured amplitude of the BBH background as shown in \cite{{Mukherjee:2020tvr}, {Buscicchio:2020cij}}.

Under the presence of a lens, the corresponding strongly lensed GW signal is magnified such that $\rho_l=\sqrt{\mu}\rho$ \citep{Wang:1996}, where $\rho_l$ and $\rho$ are the signal-to-noise ratios (SNR) under the lensed and unlensed models, respectively, and $\mu$ is the relative magnification factor \citep{Narayan:1996, Hannuksela:2019}.
Since the gravitational wave frequency evolution is not affected by strong lensing, the lensing magnification is equivalent to a scaling of the source luminosity distance by a factor of  $1/\sqrt{\mu}$ \citep{{Wang:1996}, {Dai:2016igl}}. Thus, a loud and nearby GW source could potentially be lensed and thus appear to be more distant than it seems. Also, the measured detector frame chirp mass $\mathcal{M}_c$ will typically be biased towards larger values than in the source frame since it depends directly on the redshift $z$ for the source: $\mathcal{M}_c=(1+z)\mathcal{M}_c^{source}$. Thus strong lensing results in an overestimation of the source frame masses if the signal is highly magnified. 

In addition, according to \citep{Takahashi:2003, Lai:2017arXiv170204724D}, lensing shifts the original phase of the waveform by $\Delta \phi$ in such a way that the shift is absorbed into the phase of the coalescence $\Delta \phi_c$ in the case of of gravitational waves with relation $\Delta \phi = 2 \Delta \phi_c$ (except if precession, eccentricity, or higher modes are present) \citep{Ezquiaga:2020gdt}. The shift depends on the type of lensed image: Type-I induces no phase shift, type-II induces a $+\pi/2$ phase shift, and type-III images induce a $+\pi$ phase shift. Type-III images are typically suppressed and rarely seen in the electromagnetic band, save for some rare exceptions \citep{Dahle:2012bd,Collett:2017ksf}. Therefore, one would typically expect lensed gravitational waves to consist of type-I or type-II images.

During the second observing run of Advanced LIGO and Advanced Virgo, seven binary black holes (BBH) \citep{LVC:GW170104, LVC:GW170608, LVC:GW170814, LVC:catalog} and one binary neutron star (BNS) \citep{LVC:GW170817} were detected. 
A search for gravitational-wave lensing signatures on the GWTC-1 catalog \citep{LVC:catalog} was performed in \citep{Hannuksela:2019}, but no good evidence of strong lensing was found. Note that the highest Bayes factor event pair in the analysis was the GW170104-GW170814 pair, but this was disfavored due to 1) the large time-delay between the events and 2) the prior probability of lensing being low, around $\sim 10^{-5}\, \rm yr^{-1}$ at O1 sensitivity \citep{Smith:2017mqu} for galaxy cluster lensing, and relative lensing rate $\sim 10^{-3}$ for galaxy lensing \citep{Ng:2018}. If one or more of the observed images is of type-III, the rate is understood to be significantly lower. 

The same event pair was studied in more detail in \citep{Dai:2020tpj}, which appeared at the time of writing of this article. A third, sub-threshold image consistent with the lensing hypothesis was found, GWC170620, which was first discovered in the PyCBC sub-threshold search \citep{Nitz:2019hdf}. The authors further analyzed the image configurations required for the lensing hypothesis, finding that it would consist of either one or two type-III images, and thus would require a galaxy cluster lens due to the long time delay. 
Neglecting the a priori probability of lensing, the false alarm probability for the double (triplet) was estimated at $\sim 10^{-4} - 10^{-2}$ depending on specific O2 GW events \citep{Dai:2020tpj}. 
However, when accounting for the prior probability of lensing and the fact that the observed images would require a very peculiar image configuration, the lensing hypothesis is disfavored: the authors conclude that there is not sufficient evidence to conclude that the event pair is lensed. 
The authors exclude the mass and other binary parameters that rely on the knowledge of the BBH population parameters to determine the false alarm rate of the pair due to lensing.

If the double/triplet events were lensed, then it would likely imply that the existing estimates of the lensing statistics are likely incorrect in predicting the relative fraction of galaxy cluster lenses and the total rate of lensed events (and hence the merger rate density of BBHs at high redshift). Another likely implication is a population of lenses which can form type-III images more frequently than observed in the electromagnetic spectrum. 
To reconcile for the discrepancy, one would likely require all of the following: 
1) the merger rate density of BBHs to rise at a higher rate than existing estimates from the usual formation channels, 
2) galaxy cluster lenses to make up a significant portion of the lensing optical depth, and
3) prominence of lensing configurations that can form heavily magnified type-III images in GW channels but not in electromagnetic channels. 

In this paper, we present a Bayesian model selection method similar to \cite{Haris:2018arXiv180707062H}, but instead of computing the lensing model evidence using kernel density estimation (KDE) from independent event posterior samples, we calculate the lensing evidence directly with parameter estimation by jointly fitting both images. We explicitly test the expected phase shifts and use an astrophysically motivated prior for the relative magnification factor. Moreover, we calculate the Bayes factors between the lensed and unlensed hypothesis and from the measured time delays, we determine the prior odds for any two events to be likely images of each other to produce an odds ratio that we can use to test the lensed and unlensed hypothesis.
 
\section{\label{sec:level1}Gravitational lensing model selection}

For a GW signal at luminosity distance $D_L$, the amplitude of the corresponding lensed images are magnified by a factor of $\sqrt{\mu_i}$, where $i$ labels the corresponding absolute magnification factor for each image so that the observed luminosity distances will be,
\begin{equation}
D_L^{(i)}=D_L/\sqrt{\mu_i}\,,
\end{equation}
Since the magnification factors and luminosity distance are degenerate, the individual magnification factors are difficult to constrain. Thus, we instead use the relative magnification factor $\mu$,
\begin{equation}
\mu = \left( \frac{D_L^{(1)}}{D_L^{(2)}} \right)^2 = \frac{\mu_2}{\mu_1} \,,
\end{equation}
where $D_L^{(1)}$ and $D_L^{(2)}$ are the observed luminosity distances of the first and second images respectively and $\mu_1$ and $\mu_2$ are the corresponding absolute magnification factors.  

For strong lensing, the probability distribution for the individual magnifications is well known in the high-magnification limit and is given by $p(\mu_i) \propto \mu_i^{-3}$ \citep{Blandford:1986zz}. In this work, we assume that the two magnifications are independent; however, we note that this is an approximation as the two are in practice related through the lensing model.

Given two observed detector strains $d_1(t)$ and $d_2(t)$ with confirmed GW detections, we want to determine whether these two signals are lensed or not.
The lensed hypothesis $\HL$ states that the two signals come from the same astrophysical source and are thus lensed. Meanwhile, the unlensed model $\HU$ assumes that the two signals are from independent astrophysical sources.
Under the lensing hypothesis, we first introduce a set of common parameters for the two events, $\eta = \{m_1, m_2, a_1, a_2, \iota, \alpha, \delta, \psi\}$, where $m_1$ and $m_2$ are the component source frame masses, $a_1$ and $a_2$ are the component spins, $\iota$ is the inclination angle of the binary, $\alpha$ and $\delta$ are the right ascension and declination, and $\psi$ is the polarization angle. We also introduce lensing dependent parameters, $\zeta = \{D_L, \phi_c, t_c\}$ where $D_L$ is the luminosity distance to the source, $\phi_c$ is the coalescence phase, and $t_c$ the time at coalescence. Hence for the lensed hypothesis, we expect the common parameters $\eta$ to be the same for the two events and only for the lensing dependent parameters to differ. Thus, the likelihood under the lensed hypothesis, given GW strain data $d_1$ and $d_2$, can be written as:
\begin{equation}
P(d_1, d_2|\vec{\theta_1}, \HL) = P(d_1|\eta, \zeta_1, \HL)P(d_2|\eta, \zeta_2, \HL) 
\end{equation}
where, $d_1$,  $\zeta_1$, and $d_2$,  $\zeta_2$ are the data and independent parameters for the first and second images respectively. Note, that for this model we sample the magnification factor $\mu$ instead of $D_L^{(2)}$ in the independent parameters $\zeta_2$.

For the unlensed hypothesis, the parameters of the two events are sampled independently. The likelihood in the unlensed hypothesis $\HU$, is simply the product of the likelihoods of the two events since they are independent of each other,
\begin{equation}
P(d_1, d_2|\vec{\theta_2}, \HU) = P(d_1|\eta_1, \zeta_1, \HU)P(d_2|\eta_2, \zeta_2, \HU) 
\end{equation}
where $\eta_1$, $\zeta_1$, and $\eta_2$, $\zeta_2$ are the parameters for the first and the second GW events respectively.

To compare the two models, we compute the ratio of the evidences $P(d_1, d_2 |n_j,\HL)$ and $P(d_1, d_2|\HU)$, also known as the Bayes factor,
\begin{widetext}
\begin{equation}
B^L_{U} = \frac{P(d_1, d_2|n_j, \HL)}{P(d_1, d_2|\HU)}
        = \frac{\int P(d_1|\eta, \zeta_1, \HL)P(d_2|\eta, \zeta_2, \HL) P(\eta, \zeta_1, \zeta_2|\HL,n_j) d\eta d\zeta_1 d\zeta_2}{\int P(d_1|\eta_1, \zeta_1, \HU)P(d_2|\eta_2, \zeta_2, \HU) P(\eta_1, \zeta_1, \eta_2, \zeta_2|\HU) d\eta_1 d\zeta_1 d\eta_2 d\zeta_2}
\end{equation}
\end{widetext}
where $n_j$ is the Morse index which determines the type of lensing image (type-I/II/III) and thus the expected phase difference for the pair, while $P(\eta, \zeta_1, \zeta_2|\HL,n_j)$ and $P(\eta_1, \zeta_1, \eta_2, \zeta_2|\HU)$ are the priors for the lensed and unlensed hypothesis, respectively. We calculate the lensing evidence with nested sampling \citep{skilling:2006, Veitch:2015} using \textsc{\texttt{lalinference\_nest}} \citep{LALSuite}.

The time delay between any two events can also be used to compute a corresponding timing Bayes factor \citep{Haris:2018arXiv180707062H, Hannuksela:2019},
\begin{equation}
B_t = \frac{P(\Delta t|\HL)}{P(\Delta t|\HU)}
\end{equation}
We estimate the probability distribution $P(\Delta t|\HL)$ through simulation following the methodology of \citep{Haris:2018arXiv180707062H}. We compute $P(\Delta t|\HU)$, by assuming that independent (unlensed) events are Poisson distributed. 

To obtain the odds ratio for the lensed and unlensed hypothesis we compute,
\begin{equation}
O^L_{U} = \frac{P(d_1, d_2|n_j,\HL)}{P(d_1, d_2|\HU)}\frac{P(\Delta t|\HL)}{P(\Delta t|\HU)}\frac{P(\HL)}{P(\HU)}
\label{eq:posterior_odds}
\end{equation}
where the ratio $P(\HL)/P(\HU)$ is the prior odds for lensing compared to the unlensed event model.
However, type-III images are very rare, and hence it is likely that the image shift corresponding to $\Delta \phi=\pm \pi$ is heavily disfavored: $p(n_j=1|\HL)\ll p(n_j=\{0,1/2\}|\HL)$.

The prior odds reflects our belief on the probability of lensing for any two events and can be estimated through simulations as well as from electromagnetic observations.
As such, we compute this via the ratio of expected lensed event to independent event rates. 
The relative lensed event rate has been estimated for galaxy lenses to be around $p(\HL)/p(\HU) \sim 10^{-3}$ at design sensitivity \citep{Li:2018prc,Oguri:2018muv}, while Ref. \citep{Smith:2017mqu} finds the relative rate of galaxy cluster lensing to be $p(\HL) \sim 10^{-5} {yr}^{-1}$ at O1 sensitivity; we expect this to be somewhat larger at O2 sensitivity.

When comparing two models, the Bayesian evidence penalizes a more complex model. If we compare two models, the one with smaller prior volume or fewer parameters would be favored. This penalty is known as the Occam factor \citep{Thrane:2019}, which is automatically achieved by Bayesian inference. In our analysis, the lensing model has fewer parameters due to the parameter sharing. 
Indeed, when a signal is consistent with the lensed hypothesis, the magnitude of the Bayes factor is entirely set by the prior volume; a larger prior can increase the Bayes factor by several orders of magnitude and vice versa.
In order to reduce the prior volume difference between the two models, we impose a uniform in $\log(m_1)$ and $\log(m_2)$ prior, instead of the typical uniform priors on $m_1$ and $m_2$ (both in the detector frame). We impose a prior within the mass range of $1-100 M_{\odot}$. The difference between the prior volumes can be reduced by a factor of $10^2-10^3$ when using the uniform in log space prior. We note that an astrophysically motivated mass prior could be used instead, such as the power-law model used by the LVC \citep{LIGOScientific:2018jsj}, however, for the reasons stated above we decided to use the uniform in log prior instead.

We also take care of selection effects in the joint parameter estimation since gravitational-wave detectors are not sensitive at detecting all the binaries in the prior parameter space. Thus, we incorporate a selection function in the parameter estimation directly, which keeps a sample if it is above the detection SNR threshold, otherwise it rejects the sample. We also note that taking into account selection effects is important when one or both events are below the detection threshold \citep{Li:2019osa,McIsaac:2019use}, as will be the case when performing the joint parameter estimation using the potential third image, GWC170620 \citep{Dai:2020tpj}.

\section{Results}
We analyze potential pairs of lensed events from the second observing run of Advanced LIGO and Virgo. Due to the high computational cost of the parameter estimation, we select pairs of events that have similar sky localizations. We then run \textsc{\texttt{lalinference\_nest}} \citep{LALSuite} to obtain the lensed and unlensed model evidences. We apply the selection function implemented in \textsc{\texttt{lalinference\_nest}} to the parameter estimation and set the network SNR threshold to 10 for single events and 14 for joint events, except for the sub-threshold event GWC170620.  We sample uniformly in $\log(m1)$ and $\log(m2)$ for both lensed and unlensed models in order to mitigate the prior volume difference between models. The waveform used in our analysis is the IMRPhenomD approximant\citep{Husa:2016, Khan:2016}, a non-precessing and spin aligned (22-mode only) frequency domain BBH waveform which enable us to test the different coalescence phase shifts due to different image types. Since there is no evidence that precession has been observed in any of the events detected in O1 and O2 \citep{LVC:catalog}, we expect that the IMRPhenomD model is sufficiently accurate for this analysis.

\begin{figure*}
\centering
\includegraphics[width=0.8\textwidth]{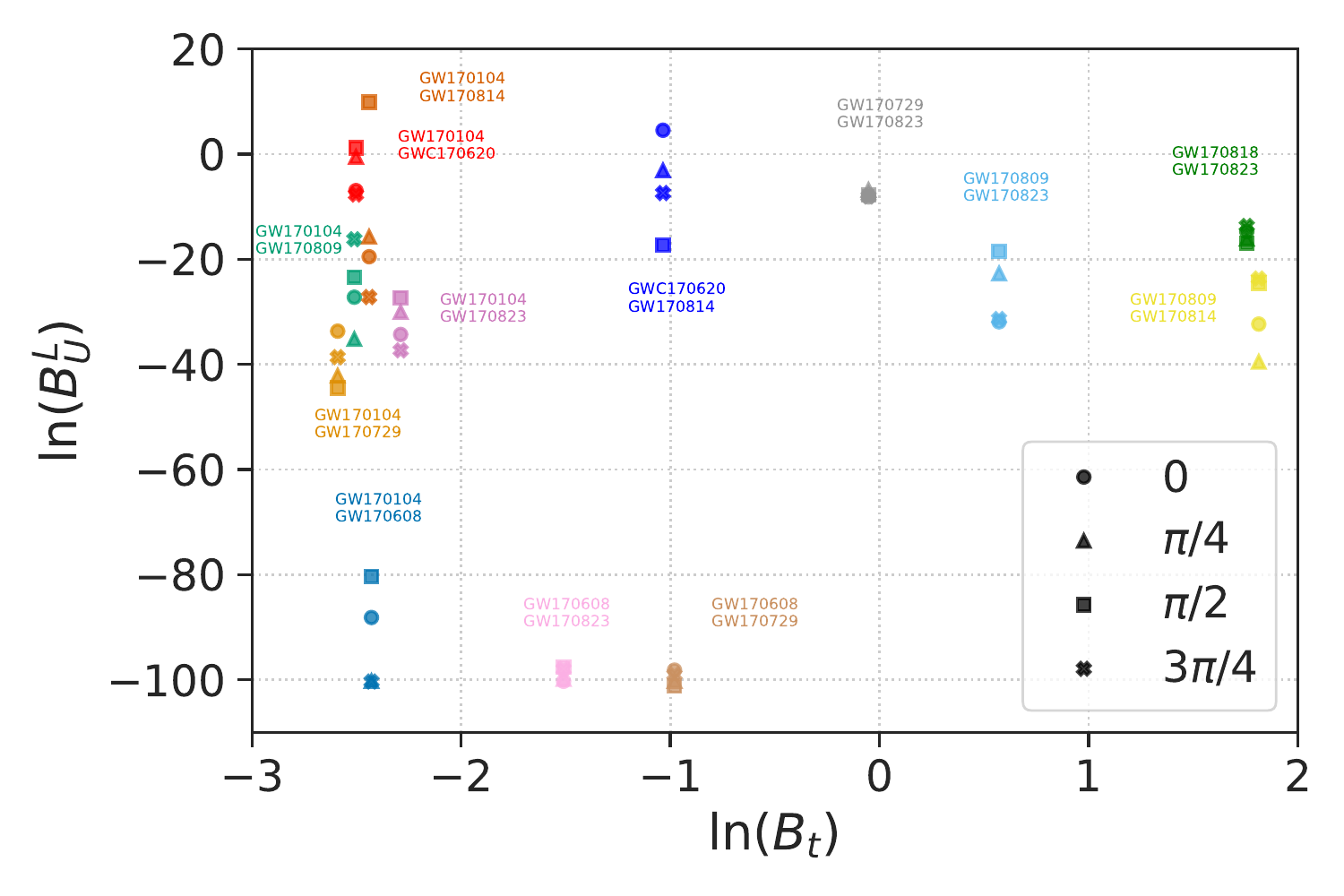}
\caption{\label {fig:BF} Natural logarithm of the Bayes factor $B^L_{U}$ and $B_t$ with $0$, $\pi/4$, $\pi/2$, and $3\pi/4$ coalescence phase shifts for pairs of events detected in O2. The Bayes factors $B^L_{U}$ are computed using \textsc{\texttt{lalinference\_nest}}, and the Bayes factor $B_t$ is computed using the time delay between any two events.}
\end{figure*}

\begin{figure*}
\centering
\includegraphics[width=0.8\textwidth]{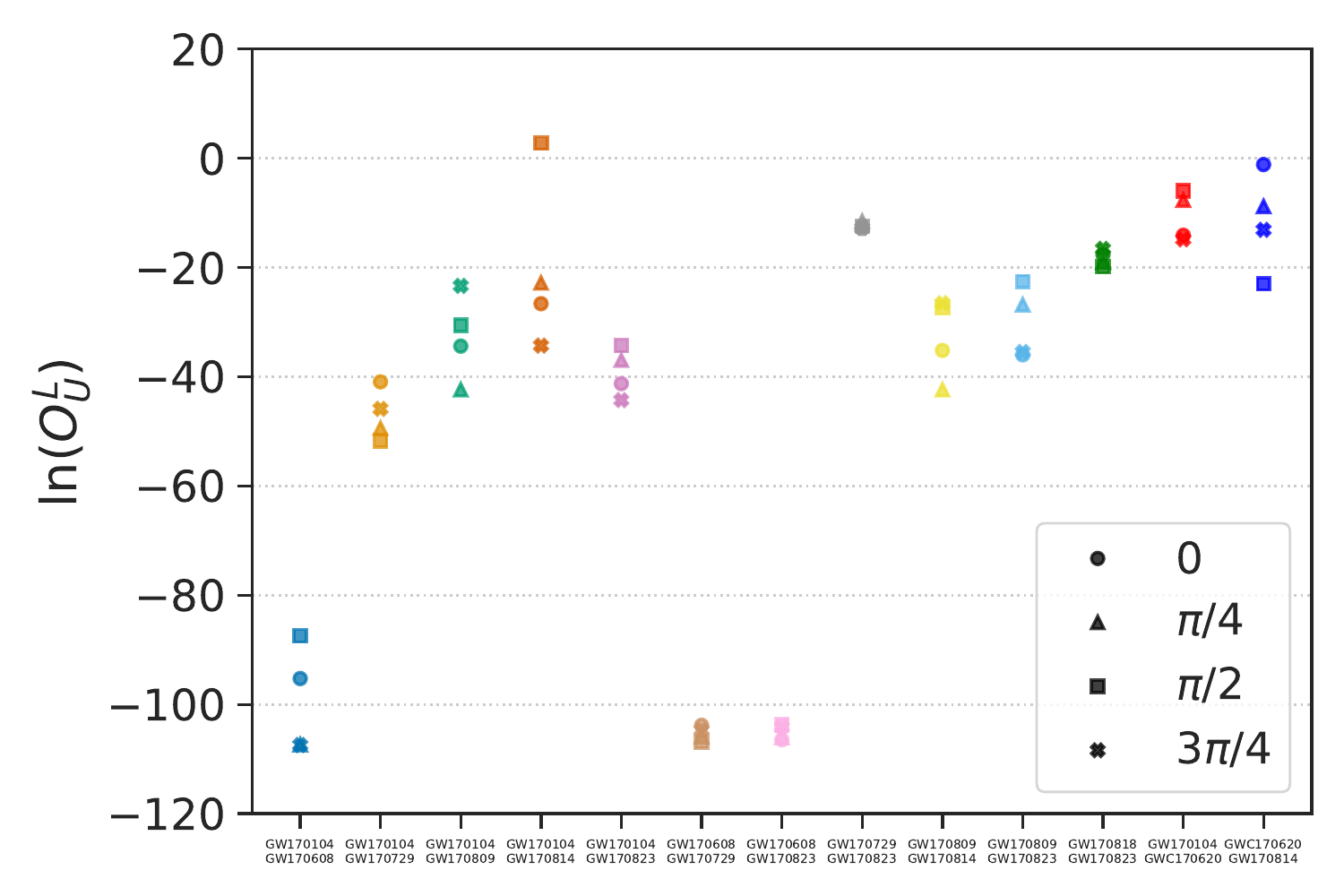}
\caption{\label {fig:PO} Natural logarithm of the odds ratio $O^L_{U}$ with $0$, $\pi/4$, $\pi/2$, and $3\pi/4$ coalescence phase shifts for pairs of events detected in O2. The odds ratios are computed using Eq. \ref{eq:posterior_odds}.}
\end{figure*}

In Fig. \ref{fig:BF} we show the Bayes factors $B^L_{U}$ and $B_t$ for pairs of events in O2. The lensed model for each pair of events are evaluated with the four different possible coalescence phase shifts. For the unlensed model, we sample the phases for each event independently. The GW170104-GW170814 pair with $\pi/2$ coalescence phase shift has the largest Bayes factor $B^L_{U}$ $\sim 1.98 \times 10^4$, favoring the lensed hypothesis in the absence of prior probability.
The event could still be an unlensed event, however, 
as the two events could be from independent sources that have similar parameters.
Therefore, we note that a high Bayes factor is not necessarily indicative of lensing. 
Nevertheless, it is intriguing that the event favours the lensing hypothesis even when including all of the binary parameters. The GW170809-GW170814 pair, which was suggested as a lensed event by Ref. \cite{Broadhurst:2019arXiv190103190B}, is clearly disfavored by the model selection. 

\begin{figure}
\centering
\includegraphics[width=0.5\textwidth]{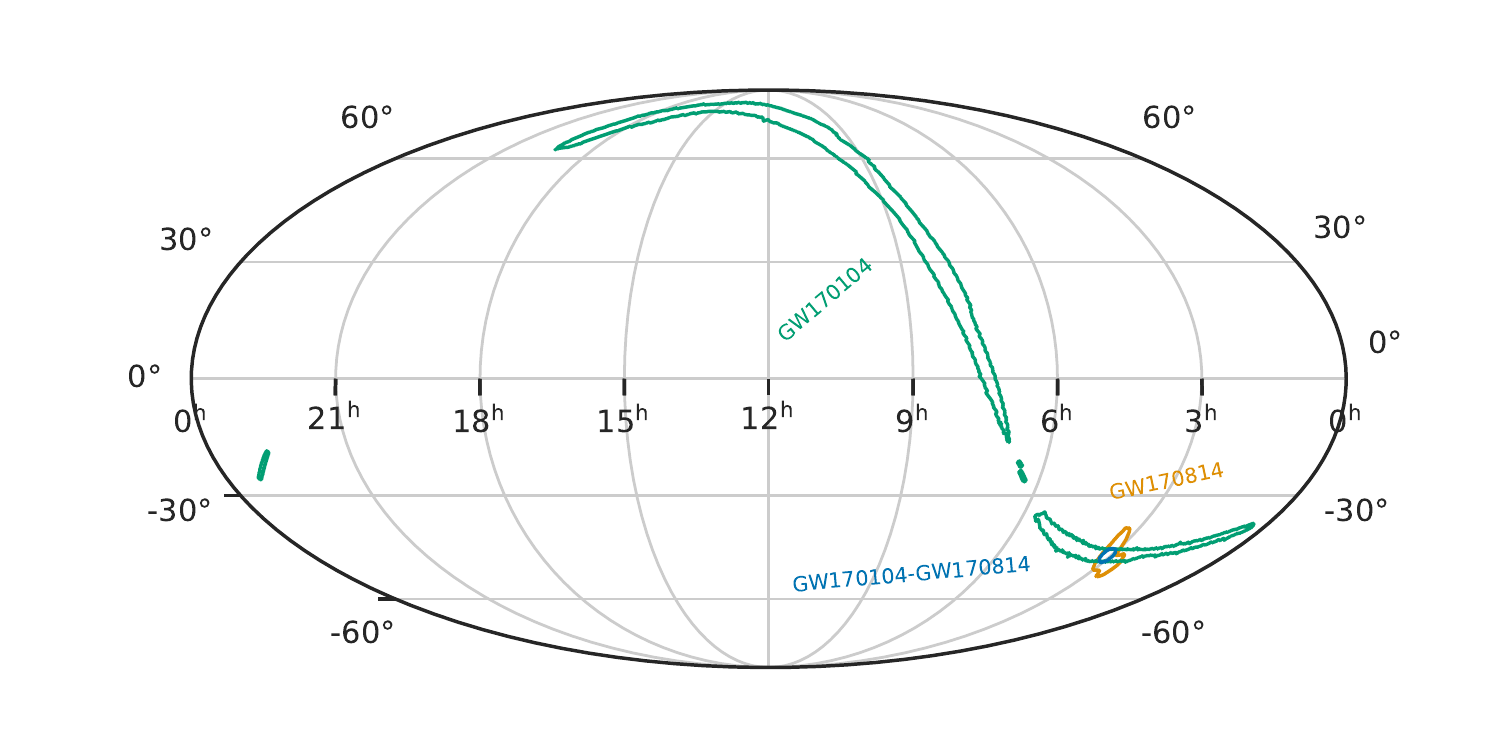}
\caption{\label {fig:skymap} The contours show 90\% confidence regions for the sky localizations posteriors of GW170104, GW170814 (treated as independent events) as well as the joint GW170104-GW170814 sky localization posterior inferred under the lensing hypothesis with a coalescence phase shift of $\pi/2$.}
\end{figure}

\begin{figure}
\centering
\includegraphics[width=0.5\textwidth]{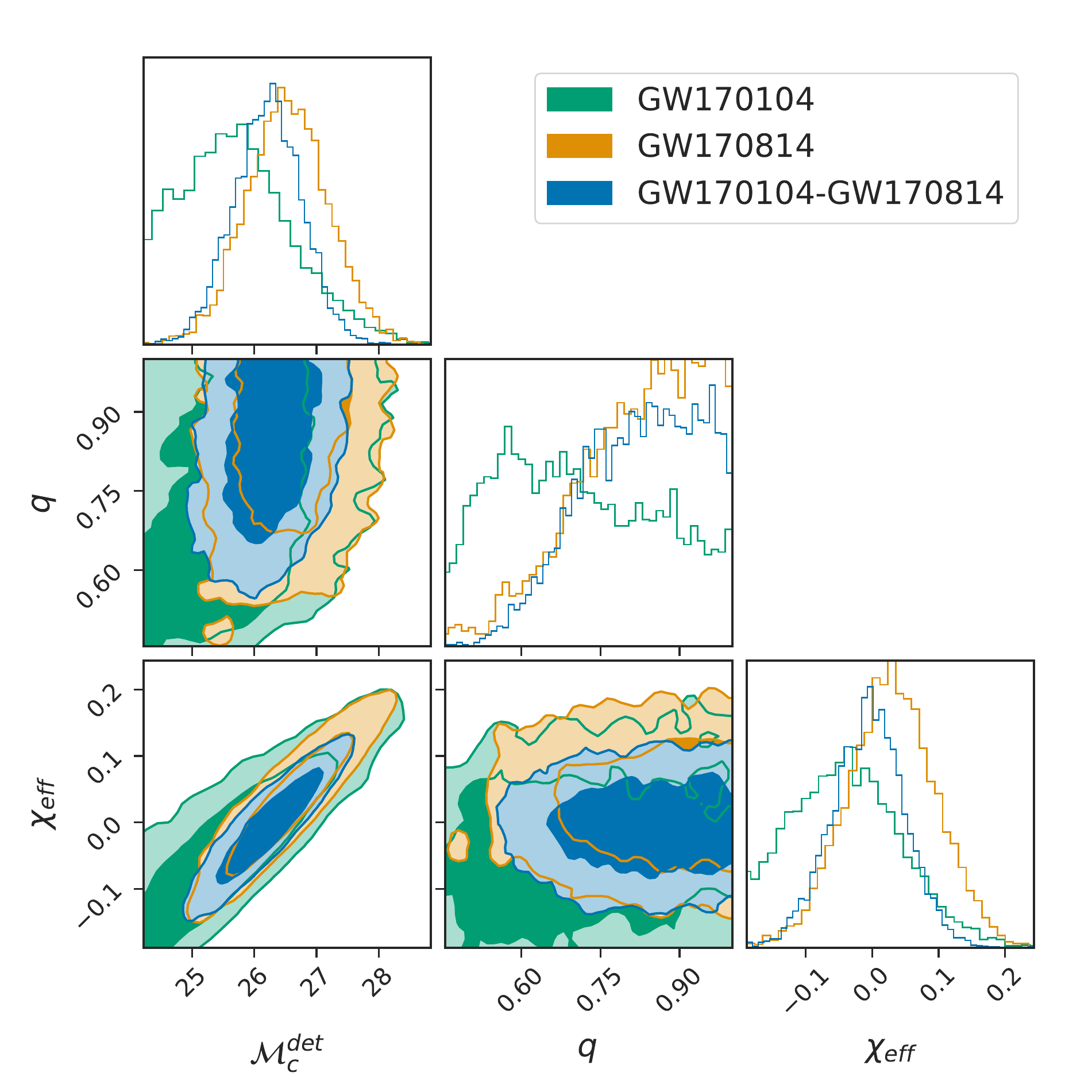}
\caption{\label {fig:corner} Corner plot showing the posterior distributions of the detector-frame chirp mass $\mathcal{M}_c$, mass ratio $q = m_2/m_1$, and the effective spin parameter $\chi_{eff}$. We show the 68\% and 98\% credible regions for the independent GW170104, GW170814 posteriors as well the jointly inferred posteriors for the GW170104-GW170814 pair under the lensing hypothesis (with coalescence phase shift of $\pi/2$).}
\end{figure}

The sky localization posterior for the GW170104-GW170814 pair is shown in Fig. \ref{fig:skymap}. The sky localization posterior inferred under the lensing hypothesis (joint parameter estimation) is better constrained and lies within the overlap region of the GW170104 and GW170814 independent parameter estimation runs. The 90 percent confidence region is better constrained because the joint run has higher SNR than each individual run and benefits from the "extra detectors" (more baselines for localization) due to the different times of arrival of each image with respect to the rotation of the earth. We also show the posterior distributions over the parameters that we expect to be unchanged due to lensing in Fig. \ref{fig:corner}. Similarly, the parameters inferred under the lensing hypothesis are better constrained compared to those inferred independently.


Assuming an observation time of 9 months for O2, and the simulations in \cite{Haris:2018arXiv180707062H} we obtain an estimate for $P(\Delta t|L)$. For the unlensed case, we assume that the detected event rate follows a Poisson distribution, that is, $P(\Delta t|U) = 2(T-\Delta t)/T^2$, where $T$ is the observation time. In Fig. \ref{fig:BF} we show the Bayes Factor $B_t$ of the lensing model with four different phase shifts compared the unlensed model. Due to the $\sim 7$ month time delay between GW170104 and GW170814, the Bayes factor $B_t$ is $\sim 8.7\times10^{-2}$. The Bayes factor $B_t$ can be large if the time delay is only a few days, such as for the GW170809-GW170814 and GW170818-GW170823 pairs, which results in $B_t \sim 6$.

The two Bayes factors $B^L_{U}$ and $B_t$ together with the prior odds for lensing can be combined to compute the odds ratio (Eq. \ref{eq:posterior_odds}). The lensed event rate is estimated for O2 to be $\sim 0.1 \text{yr}^{-1}$ \citep{Ng:2018, Hannuksela:2019}. During O2, there were eight events detected in nine months of observing time, therefore the prior odds are $\sim 0.009$. In Fig. \ref{fig:PO}, we show the odds ratio for events detected in O2. The odds ratio for GW170104 and GW170814 with $\pi/2$ coalescence phase shift is $O^L_{U} \sim 20$. It is the only pair of events that moderately prefers the lensing hypothesis even with prior information folded in. For other pairs, detected in O2, we do not see any lensed evidence as the odds ratios are much less than 1.

Our $B_t$ and prior odds estimations are based on galaxy lensing. However, the long time delay may point to lensing by a galaxy cluster. Galaxy cluster lensing is expected to be rare at O2 sensitivity. The prior probability of galaxy cluster lensing is $\sim 10^{-5} \text{yr}^{-1}$ \citep{Smith:2017mqu}.
Meanwhile, because the phase shift corresponds to $\pi$, or a type-III image as pointed in \cite{Dai:2020tpj}, the probability of lensing is further disfavored: The probability of observing type-III images should be very low, such that $p(n_j|\HL) \ll 1$. Indeed, type-III images are rarely observed in the electromagnetic band.

\section{Discussion and Conclusions}

To summarize, we used Bayesian model selection to identify lensed events in the second observing run of Advanced LIGO and Virgo. We test the lensing model with $0, \pi/4$, $\pi/2$, and $3\pi/4$ coalescence phase shifts and take care of selection effects. 
The most significant event pair is found to be GW170104-GW170814 at a very high Bayes factor; however, 
the two signals may come from two independent sources and have similar parameters and therefore it is not clear if this is indicative of lensing. 
Moreover, the event is disfavored as a lensed candidate based on the understanding of lens configurations and the BBH and lens populations by an overwhelming amount: Indeed, the probability of observing lensing configurations with these type of time-delays and image configurations is 
low for both galaxies and galaxy clusters. 
The estimates of lensed rates and the relative contribution of galaxies and galaxy clusters may vary to some degree, but no current estimate predicts that galaxy cluster lensing should become prominent at O2 sensitivity. 

Let us then entertain the possibility that the event was lensed. If this were the case, then unless we were to accept that we were simply incredibly lucky, it would likely imply all of the following: 
\begin{enumerate}
 \item The relative contribution of galaxy cluster lensing is more important than previously believed, which would explain the high time-delay between the events. Refs. \cite{Smith:2017mqu,Smith:2018gle,Smith:2018kbc,Robertson:2020mfh} have studied galaxy cluster lensing, and argued that highly magnified events have been historically observed more prominently lensed by galaxy cluster scale lenses.
 \item The merger-rate density of binary black holes must rise at a significantly higher rate than previously predicted. Indeed, the current lensing rate estimates rely on black holes tracing the star-formation rate density. For example, the Belczynski distribution is often used to model the merger-rate density \citep{Belczynski:2016obo}. Out of the studied models of black hole formation, none predict high enough merger rates that galaxy cluster lensing would become observable at O2 sensitivity \citep{Oguri:2018muv}.
 \item Type-III images are more prominent for gravitational-wave sources than they are for electromagnetic sources. 
\end{enumerate}

Let us therefore state that extraordinary claims require extraordinary evidence. Based on the prior probability of lensing by these type of systems, we advise the reader to be very careful in interpreting the results: There is no sufficient evidence to claim that the event is lensed. Indeed, in the absence of clear-cut evidence to the contrary, we must conclude that there is no sufficient evidence to claim that the event is lensed, in agreement with \citep{Dai:2020tpj} and \citep{Hannuksela:2019}. 

However, two pieces of evidence could together possibly determine if the events were lensed. First, it is vital to perform injection campaigns to determine the probability of a non-lensed event. A similar study was conducted in \citep{Dai:2020tpj} with false alarm probability between $10^{-4}$ and $10^{-2}$ for O2 events. 
Moreover, another intriguing possibility of cross-verification is through searches in the electromagnetic channels, as pointed out by \citep{Dai:2020tpj}: If the events are lensed, then their host galaxy must also be lensed \citep{Hannuksela:2020xor}. If the third event proposed as a lensed candidate for the pair in \citep{Dai:2020tpj}, then it would likely afford us three time delays, which would allow for a unique opportunity to localize the host galaxy and the galaxy cluster which lensed it in an electromagnetic follow up. Ref. \citep{Hannuksela:2020xor} demonstrated that such a search is possible for galaxies. We note that due to the rarity of galaxy clusters, the search is expected to be even more powerful for galaxy clusters. In the case of doubly lensed events such as the GW170104-GW170814 pair, the single time-delay estimate may be quite degenerate with the lens parameters and the source alignment.

\section{Acknowledgments}
The authors would like to thank Otto Hannuksela, K. Haris, John Veitch, Rico K. L. Lo, Ajit Mehta, Jose Ezquiaga, Daniel Holz, David Keitel, Suvodip Mukherjee, Ken K. Y. Ng, Leo Singer, and Will Farr for useful discussions. IMH is supported by the NSF Graduate Research Fellowship Program under grant DGE-17247915. This work was supported by NSF awards PHY-1607585 and PHY-1912649. The authors are grateful for computational resources provided by the LIGO Laboratory and supported by National Science Foundation Grants PHY-0757058 and PHY-0823459, and those provided by the Leonard E Parker Center for Gravitation, Cosmology and Astrophysics at the University of Wisconsin-Milwaukee. We thank LIGO and Virgo Collaboration for providing the data for this work. This article has been assigned LIGO document number LIGO-P1900017.

\bibliography{references}

\begin{thebibliography}{}
\expandafter\ifx\csname natexlab\endcsname\relax\def\natexlab#1{#1}\fi
\providecommand{\url}[1]{\href{#1}{#1}}
\providecommand{\dodoi}[1]{doi:~\href{http://doi.org/#1}{\nolinkurl{#1}}}
\providecommand{\doeprint}[1]{\href{http://ascl.net/#1}{\nolinkurl{http://ascl.net/#1}}}
\providecommand{\doarXiv}[1]{\href{https://arxiv.org/abs/#1}{\nolinkurl{https://arxiv.org/abs/#1}}}

\bibitem[{Aasi {et~al.}(2015)}]{TheLIGOScientific:2014jea}
Aasi, J., {et~al.} 2015, Class. Quant. Grav., 32, 074001,
  \dodoi{10.1088/0264-9381/32/7/074001}

\bibitem[{Abbott {et~al.}(2018)}]{Aasi:2013wya}
Abbott, B., {et~al.} 2018, Living Rev. Rel., 21, 3,
  \dodoi{10.1007/s41114-018-0012-9}

\bibitem[{Abbott {et~al.}(2019)}]{LIGOScientific:2018jsj}
---. 2019, Astrophys. J. Lett., 882, L24, \dodoi{10.3847/2041-8213/ab3800}

\bibitem[{{Abbott} {et~al.}(2019){Abbott}, {Abbott}, {Abbott}, {Abraham}, \&
  {Virgo Collaboration}}]{LVC:catalog}
{Abbott}, B.~P., {Abbott}, R., {Abbott}, T.~D., {Abraham}, S., \& {Virgo
  Collaboration}, L. 2019, Physical Review X, 9, 031040,
  \dodoi{10.1103/PhysRevX.9.031040}

\bibitem[{{Abbott} {et~al.}(2017{\natexlab{a}}){Abbott}, {Abbott}, {Abbott},
  {Acernese}, {LIGO Scientific}, \& {Virgo Collaboration}}]{LVC:GW170104}
{Abbott}, B.~P., {Abbott}, R., {Abbott}, T.~D., {et~al.} 2017{\natexlab{a}},
  \prl, 118, 221101, \dodoi{10.1103/PhysRevLett.118.221101}

\bibitem[{{Abbott} {et~al.}(2017{\natexlab{b}}){Abbott}, {Abbott}, {Abbott},
  {Acernese}, {(LIGO Scientific Collaboration}, \& {Virgo
  Collaboration}}]{LVC:GW170608}
---. 2017{\natexlab{b}}, \apjl, 851, L35, \dodoi{10.3847/2041-8213/aa9f0c}

\bibitem[{{Abbott} {et~al.}(2017{\natexlab{c}}){Abbott}, {Abbott}, {Abbott},
  {Acernese}, {LIGO Scientific Collaboration}, \& {Virgo
  Collaboration}}]{LVC:GW170814}
---. 2017{\natexlab{c}}, \prl, 119, 141101,
  \dodoi{10.1103/PhysRevLett.119.141101}

\bibitem[{{Abbott} {et~al.}(2017{\natexlab{d}}){Abbott}, {Abbott}, {Abbott},
  {Acernese}, {LIGO Scientific Collaboration}, \& {Virgo
  Collaboration}}]{LVC:GW170817}
---. 2017{\natexlab{d}}, \prl, 119, 161101,
  \dodoi{10.1103/PhysRevLett.119.161101}

\bibitem[{Acernese {et~al.}(2015)}]{TheVirgo:2014hva}
Acernese, F., {et~al.} 2015, Class. Quant. Grav., 32, 024001,
  \dodoi{10.1088/0264-9381/32/2/024001}

\bibitem[{Belczynski {et~al.}(2016)Belczynski, Holz, Bulik, \&
  O'Shaughnessy}]{Belczynski:2016obo}
Belczynski, K., Holz, D.~E., Bulik, T., \& O'Shaughnessy, R. 2016, Nature, 534,
  512, \dodoi{10.1038/nature18322}

\bibitem[{Blandford \& Narayan(1986)}]{Blandford:1986zz}
Blandford, R., \& Narayan, R. 1986, Astrophys. J., 310, 568,
  \dodoi{10.1086/164709}

\bibitem[{{Broadhurst} {et~al.}(2019){Broadhurst}, {Diego}, \&
  {Smoot}}]{Broadhurst:2019arXiv190103190B}
{Broadhurst}, T., {Diego}, J.~M., \& {Smoot}, George~F., I. 2019, arXiv
  e-prints, arXiv:1901.03190.
\newblock \doarXiv{1901.03190}

\bibitem[{Buscicchio {et~al.}(2020)Buscicchio, Moore, Pratten, Schmidt,
  Bianconi, \& Vecchio}]{Buscicchio:2020cij}
Buscicchio, R., Moore, C.~J., Pratten, G., {et~al.} 2020.
\newblock \doarXiv{2006.04516}

\bibitem[{Collett {et~al.}(2017)}]{Collett:2017ksf}
Collett, T.~E., {et~al.} 2017, Astrophys. J., 843, 148,
  \dodoi{10.3847/1538-4357/aa76e6}

\bibitem[{Dahle {et~al.}(2013)}]{Dahle:2012bd}
Dahle, H., {et~al.} 2013, Astrophys. J., 773, 146,
  \dodoi{10.1088/0004-637X/773/2/146}

\bibitem[{{Dai} \& {Venumadhav}(2017)}]{Lai:2017arXiv170204724D}
{Dai}, L., \& {Venumadhav}, T. 2017, arXiv e-prints, arXiv:1702.04724.
\newblock \doarXiv{1702.04724}

\bibitem[{Dai {et~al.}(2017)Dai, Venumadhav, \& Sigurdson}]{Dai:2016igl}
Dai, L., Venumadhav, T., \& Sigurdson, K. 2017, Phys. Rev. D, 95, 044011,
  \dodoi{10.1103/PhysRevD.95.044011}

\bibitem[{Dai {et~al.}(2020)Dai, Zackay, Venumadhav, Roulet, \&
  Zaldarriaga}]{Dai:2020tpj}
Dai, L., Zackay, B., Venumadhav, T., Roulet, J., \& Zaldarriaga, M. 2020.
\newblock \doarXiv{2007.12709}

\bibitem[{Ezquiaga {et~al.}(2020)Ezquiaga, Holz, Hu, Lagos, \&
  Wald}]{Ezquiaga:2020gdt}
Ezquiaga, J.~M., Holz, D.~E., Hu, W., Lagos, M., \& Wald, R.~M. 2020.
\newblock \doarXiv{2008.12814}

\bibitem[{Hannuksela {et~al.}(2020)Hannuksela, Collett, Çal\i~\c skan, \&
  Li}]{Hannuksela:2020xor}
Hannuksela, O.~A., Collett, T.~E., Çal\i~\c skan, M., \& Li, T.~G. 2020.
\newblock \doarXiv{2004.13811}

\bibitem[{{Hannuksela} {et~al.}(2019){Hannuksela}, {Haris}, {Ng}, {Kumar},
  {Mehta}, {Keitel}, {Li}, \& {Ajith}}]{Hannuksela:2019}
{Hannuksela}, O.~A., {Haris}, K., {Ng}, K.~K.~Y., {et~al.} 2019, \apjl, 874,
  L2, \dodoi{10.3847/2041-8213/ab0c0f}

\bibitem[{{Haris} {et~al.}(2018){Haris}, {Mehta}, {Kumar}, {Venumadhav}, \&
  {Ajith}}]{Haris:2018arXiv180707062H}
{Haris}, K., {Mehta}, A.~K., {Kumar}, S., {Venumadhav}, T., \& {Ajith}, P.
  2018, arXiv e-prints, arXiv:1807.07062.
\newblock \doarXiv{1807.07062}

\bibitem[{Harry(2010)}]{Harry:2010zz}
Harry, G.~M. 2010, Class. Quant. Grav., 27, 084006,
  \dodoi{10.1088/0264-9381/27/8/084006}

\bibitem[{{Husa} {et~al.}(2016){Husa}, {Khan}, {Hannam}, {P{\"u}rrer}, {Ohme},
  {Forteza}, \& {Boh{\'e}}}]{Husa:2016}
{Husa}, S., {Khan}, S., {Hannam}, M., {et~al.} 2016, \prd, 93, 044006,
  \dodoi{10.1103/PhysRevD.93.044006}

\bibitem[{{Khan} {et~al.}(2016){Khan}, {Husa}, {Hannam}, {Ohme}, {P{\"u}rrer},
  {Forteza}, \& {Boh{\'e}}}]{Khan:2016}
{Khan}, S., {Husa}, S., {Hannam}, M., {et~al.} 2016, \prd, 93, 044007,
  \dodoi{10.1103/PhysRevD.93.044007}

\bibitem[{Li {et~al.}(2019)Li, Lo, Sachdev, Chan, Lin, Li, \&
  Weinstein}]{Li:2019osa}
Li, A.~K., Lo, R.~K., Sachdev, S., {et~al.} 2019.
\newblock \doarXiv{1904.06020}

\bibitem[{Li {et~al.}(2018)Li, Mao, Zhao, \& Lu}]{Li:2018prc}
Li, S.-S., Mao, S., Zhao, Y., \& Lu, Y. 2018, Mon. Not. Roy. Astron. Soc., 476,
  2220, \dodoi{10.1093/mnras/sty411}

\bibitem[{{LIGO Scientific Collaboration}(2018)}]{LALSuite}
{LIGO Scientific Collaboration}. 2018, {LIGO Algorithm Library},
  \dodoi{10.7935/GT1W-FZ16}

\bibitem[{McIsaac {et~al.}(2019)McIsaac, Keitel, Collett, Harry, Mozzon, Edy,
  \& Bacon}]{McIsaac:2019use}
McIsaac, C., Keitel, D., Collett, T., {et~al.} 2019.
\newblock \doarXiv{1912.05389}

\bibitem[{Mukherjee {et~al.}(2020)Mukherjee, Broadhurst, Diego, Silk, \&
  Smoot}]{Mukherjee:2020tvr}
Mukherjee, S., Broadhurst, T., Diego, J.~M., Silk, J., \& Smoot, G.~F. 2020.
\newblock \doarXiv{2006.03064}

\bibitem[{{Narayan} \& {Bartelmann}(1996)}]{Narayan:1996}
{Narayan}, R., \& {Bartelmann}, M. 1996, arXiv e-prints, astro.
\newblock \doarXiv{astro-ph/9606001}

\bibitem[{{Ng} {et~al.}(2018){Ng}, {Wong}, {Broadhurst}, \& {Li}}]{Ng:2018}
{Ng}, K. K.~Y., {Wong}, K. W.~K., {Broadhurst}, T., \& {Li}, T. G.~F. 2018,
  \prd, 97, 023012, \dodoi{10.1103/PhysRevD.97.023012}

\bibitem[{Nitz {et~al.}(2019)Nitz, Dent, Davies, Kumar, Capano, Harry, Mozzon,
  Nuttall, Lundgren, \& Tápai}]{Nitz:2019hdf}
Nitz, A.~H., Dent, T., Davies, G.~S., {et~al.} 2019, Astrophys. J., 891, 123,
  \dodoi{10.3847/1538-4357/ab733f}

\bibitem[{Oguri(2018)}]{Oguri:2018muv}
Oguri, M. 2018, Mon. Not. Roy. Astron. Soc., 480, 3842,
  \dodoi{10.1093/mnras/sty2145}

\bibitem[{Robertson {et~al.}(2020)Robertson, Smith, Massey, Eke, Jauzac,
  Bianconi, \& Ryczanowski}]{Robertson:2020mfh}
Robertson, A., Smith, G.~P., Massey, R., {et~al.} 2020,
  \dodoi{10.1093/mnras/staa1429}

\bibitem[{Skilling(2006)}]{skilling:2006}
Skilling, J. 2006, Bayesian Anal., 1, 833, \dodoi{10.1214/06-BA127}

\bibitem[{Smith {et~al.}(2017)}]{Smith:2018gle}
Smith, G., {et~al.} 2017, IAU Symp., 338, 98, \dodoi{10.1017/S1743921318003757}

\bibitem[{Smith {et~al.}(2019)Smith, Bianconi, Jauzac, Richard, Robertson,
  Berry, Massey, Sharon, Farr, \& Veitch}]{Smith:2018kbc}
Smith, G., Bianconi, M., Jauzac, M., {et~al.} 2019, Mon. Not. Roy. Astron.
  Soc., 485, 5180, \dodoi{10.1093/mnras/stz675}

\bibitem[{Smith {et~al.}(2018)Smith, Jauzac, Veitch, Farr, Massey, \&
  Richard}]{Smith:2017mqu}
Smith, G.~P., Jauzac, M., Veitch, J., {et~al.} 2018, Mon. Not. Roy. Astron.
  Soc., 475, 3823, \dodoi{10.1093/mnras/sty031}

\bibitem[{{Takahashi} \& {Nakamura}(2003)}]{Takahashi:2003}
{Takahashi}, R., \& {Nakamura}, T. 2003, \apjl, 595, arXiv:astro.
\newblock \doarXiv{astro-ph/0305055}

\bibitem[{{Thrane} \& {Talbot}(2019)}]{Thrane:2019}
{Thrane}, E., \& {Talbot}, C. 2019, \pasa, 36, e010,
  \dodoi{10.1017/pasa.2019.2}

\bibitem[{{Veitch} {et~al.}(2015){Veitch}, {Raymond}, {Farr}, {Farr}, {Graff},
  {Vitale}, {Aylott}, {Blackburn}, {Christensen}, {Coughlin}, {Del Pozzo},
  {Feroz}, {Gair}, {Haster}, {Kalogera}, {Littenberg}, {Mandel},
  {O'Shaughnessy}, {Pitkin}, {Rodriguez}, {R{\"o}ver}, {Sidery}, {Smith}, {Van
  Der Sluys}, {Vecchio}, {Vousden}, \& {Wade}}]{Veitch:2015}
{Veitch}, J., {Raymond}, V., {Farr}, B., {et~al.} 2015, \prd, 91, 042003,
  \dodoi{10.1103/PhysRevD.91.042003}

\bibitem[{{Wang} {et~al.}(1996){Wang}, {Stebbins}, \& {Turner}}]{Wang:1996}
{Wang}, Y., {Stebbins}, A., \& {Turner}, E.~L. 1996, \prl, 77, 2875,
  \dodoi{10.1103/PhysRevLett.77.2875}

\end{thebibliography}

\end{document}